\documentclass{article}
\usepackage{spconf,amsmath,graphicx}
\usepackage{amsfonts}
\usepackage{epsfig,amssymb,latexsym,graphics,float}
\usepackage{setspace,subfig}
\usepackage{citesort}
\usepackage{amsopn}
\usepackage{amssymb,times}
\usepackage[ruled,vlined,linesnumbered]{algorithm2e}
\usepackage{ifpdf}
\usepackage{dsfont}

\title{iTrace$:$ An Implicit Trust Inference Method for Trust-aware Collaborative Filtering}
\name{Xu He, Bin Liu$^{\ast}$ and Ke-Jia Chen
\thanks{$^\star$Correspondence author: Bin Liu (Email: bins@ieee.org). This work was partly supported by the National Natural Science Foundation
(NSF) of China under grant No. 61571238, China Postdoctoral Science Foundation under grant Nos. 2015M580455 and 2016T90483,
the Six Talents Peak
Foundation of Jiangsu Province under grant No. XYDXXJS-CXTD-006 and the
Scientific and Technological Support Project (Society) of Jiangsu Province under
grant No. BE2016776.
}}
\address{School of Computer Science\\
Jiangsu Key Laboratory of Big Data Security $\&$ Intelligent Processing\\
Nanjing University of Posts and Telecommunications\\
Nanjing, Jiangsu 210023, China}
\begin{document}
\maketitle
\begin{abstract}
The growth of Internet commerce has stimulated the use of collaborative filtering (CF) algorithms as recommender systems.
A CF algorithm recommends items of interest to the target user by leveraging the votes given by other similar users.
In a standard CF framework, it is assumed that the credibility of every voting user is exactly the same with respect to the
target user. This assumption is not satisfied and may lead to misleading recommendations in practice. A natural countermeasure is to
design a trust-aware CF algorithm, which can take account of the difference in the credibilities of the voting users when performing CF.
To this end, this paper presents a trust inference approach, which can predict the \emph{implicit trust} of the target user on every voting
user from a sparse \emph{explicit trust} matrix. Then an improved CF algorithm termed \emph{iTrace} is proposed,
which employs both the explicit and the predicted \emph{implicit trust} to provide recommendations.
An empirical evaluation on a public dataset demonstrates that the proposed algorithm provides a significant improvement in
recommendation quality in terms of mean absolute error.
\end{abstract}
\keywords recommender systems; trust-aware collaborative filtering; implicit trust; explicit trust; trust inference; shortest path
\section{Introduction}
With the massive growth of the internet and the emergence of electronic commerce over the last decades, recommender system (RecSys)
has become an indispensable
technique to mitigate the problem of information overload for users. The aim of RecSys is to provide target users with high quality,
personalized
recommendations, and to help them find items (e.g., books, movies, news, music, etc.) of interest from a plethora of available
choices \cite{resnick1997recommender}.

Collaborative filtering (CF) seems to be one of the most well-known and commonly used techniques to build a
RecSys \cite{sarwar2001item,herlocker2004evaluating,koren2015advances}.
The underlying idea of CF is that users with similar preferences in the past are likely to favor the same items
(e.g., books, movies, news, music, etc.) in the future. The CF method is easy to implement. A typical CF
method predicts the rating value user $u$ gives to item $i$ as follows \cite{sarwar2001item}:
\begin{equation}\label{eqn:cf}
r_{u,i}=\bar{r}_u+\frac{\sum_{v\in U}w(u,v)(r_{v,i}-\bar{r}_{v})}{\sum_{v\in U}|w(u,v)|},
\end{equation}
where $U$ denotes a set of $K$ neighbors of $u$ who rated item $i$ (also called $u'$s voting users in what follows),
$\bar{r}_u$ the average rating of user $u$ for all the items rated by $u$, and $w(u,v)$ the weight assigned to user $v'$s vote when she recommends items to
$u$. In a standard CF framework, the weight $w(u,v)$ is set as a similarity measure between users $u$ and $v$, denoted by $\mbox{sim}(u,v)$, and the
neighbors of $u$ are those most similar to $u$ who co-rated item $i$ with $u$.  In order to compute the similarity between users,
a variety of similarity measures have been proposed, such as Pearson correlation, cosine vector similarity, Spearman correlation, entropy-based
uncertainty, and mean-square difference. It is reported that Pearson correlation performs better than the
others \cite{herlocker1999algorithmic,breese1998empirical}. The \emph{Pearson correlation coefficient} is used here, defined as
follows \cite{herlocker2004evaluating,wang2006unifying}
\begin{equation}\label{eqn:pearson}
\mbox{sim}(u,v)=\frac{\sum_{i\in I}(r_{u,i}-\bar{r}_{u})(r_{v,i}-\bar{r}_{v})}{\sqrt{\sum_{i\in I}(r_{u,i}-\bar{r}_{u})^2}\sqrt{\sum_{i\in I}(r_{v,i}-\bar{r}_{v})^2}},
\end{equation}
where $I$ denotes the set of items that users $u$ and $v$ have co-rated.

In practical applications, users in general only rate a small portion of items, but accurate recommendations are expected for the
cold users who rate only a few items. This raises two inherent obstacles to obtain satisfactory
recommending quality, namely data sparsity and cold start \cite{guo2014merging,massa2004trust,massa2007trust,ricci2015recommender}.
In principle, this is caused by the lack of sufficient and reliable elements in $U$ and/or $I$ to calculate Eqns.(\ref{eqn:cf}) and (\ref{eqn:pearson}).
A possible solution to get around this is to incorporate trust relationships into the CF framework, resulting in the
trust based or trust-aware CF (TaCF) \cite{massa2004trust,guo2014merging,o2005trust,dubois2009improving,massa2007trust,ma2014improving}.
The underlying intuition supporting the working of trust-aware recommender systems (TaRS) is that
users often accept advice from trustworthy friends in real life on topics they are not expert in. So it is reasonable to expect that
considering trust relationship among users may bring in benefits in generating recommendations. Furthermore, trust can be propagated
over a network of users, hence TaRS can overcome the data sparsity and cold start problems, from which traditional CF methods suffer,
at least in concept.

A practical issue to be considered when designing a TaCF algorithm is that the explicit trust information is usually much more sparse than the users'
ratings. A trust propagation model along with an effective implicit trust inference method is desirable to overcome the above limitation. To this end,
this paper presents an applicable implicit trust inference method, based on which an improved CF algorithm
termed \emph{iTrace} (i.e., Implicit TRust-Aware Collaborative filtEring)
is proposed.
\section{The Proposed \emph{iTrace} Algorithm}\label{sec:itrace}
In this section, we present the \emph{iTrace} algorithm followed by an analysis of its connections to existent related work.
\begin{figure}[t]
\centering
\includegraphics[width=3.3in,height=2.1in]{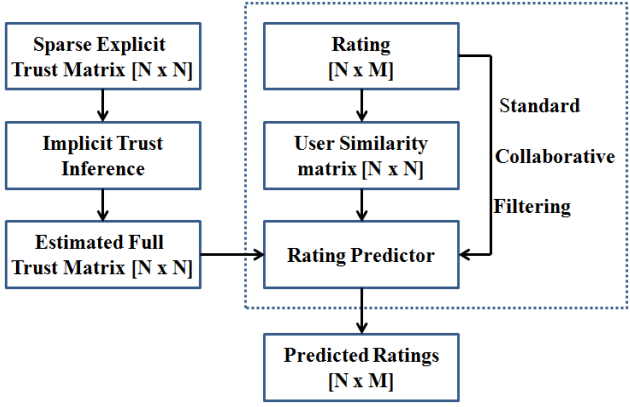}
\caption{Architecture of the \emph{iTrace} algorithm. The inputs of the algorithm include an $N\times N$ explicit trust matrix and an $N\times M$ rating matrix.
$N$ and $M$ denote the number of the users and of the items, respectively.}\label{fig:itrace_architect}
\end{figure}
\subsection{Algorithm Design}
An architecture of the \emph{iTrace} algorithm is shown in Fig.\ref{fig:itrace_architect}.
The inputs include an $N\times N$ explicit trust matrix and an $N\times M$ rating matrix, denoted in what follows by $T_e$ and $R$,
respectively. $N$ and $M$ denote the numbers of the users and of the items, respectively.
In contrast with a standard CF algorithm, \emph{iTrace} leverages much more
information except user similarity for prediction of user rating.
Such additional information is represented by a trust matrix, denoted by $\hat{T}$ in what follows, which is estimated by an implicit trust inference
module that takes $T_e$ as input. The details on the implicit trust inference module are presented in Sec.\ref{subsec:trust_inf}.
A working flow of the \emph{iTrace} algorithm for predicting user $u$'s rating value on item $i$ is summarized as follows.
\begin{enumerate}
\item Calculate the similarity metrics between $u$ and the other users who rated $i$ using Eqn.(\ref{eqn:pearson}).
\item Select the top $K$ users who are most similar to $u$ as $u$'s voting users.
\item Estimate the trust of $u$ on every voting user, using the implicit trust inference method presented in Sec. \ref{subsec:trust_inf}.
\item Predict user $u$'s rating value on item $i$ as follows
\begin{equation}\label{eqn:tacf}
r_{u,i}=\bar{r}_u+\frac{\sum_{v\in U}f(\mbox{sim}(u,v),\hat{t}(u,v))(r_{v,i}-\bar{r}_{v})}{\sum_{v\in U}|f(\mbox{sim}(u,v),\hat{t}(u,v))|},
\end{equation}
where $\hat{t}(u,v)$ is the estimated trust of $u$ on $v$, obtained from Step 3. The function $f$
plays a role of integrating user similarity and trust in rating
prediction.
\end{enumerate}

We consider two different forms of the function $f$ in our algorithm. The first one, termed incremental weighting (IW) here, is specified as follows
\begin{equation}\label{eqn:f1}
f(\mbox{sim}(u,v),\hat{t}(u,v))=\frac{\mbox{sim}(u,v)\hat{t}(u,v)}{\sum_{j\in U}\mbox{sim}(u,j)\hat{t}(u,j)}.
\end{equation}
The standard CF framework corresponds to a special case in which $\hat{t}(u,i)=\hat{t}(u,j)$ for $\forall i,j\in U$.
The other form of $f$ under consideration, termed linear weighting (LW) here, is
\begin{equation}\label{eqn:f2}
f(\mbox{sim}(u,v),\hat{t}(u,v))=\frac{\alpha \mbox{sim}(u,v)}{\sum_{j\in U}\mbox{sim}(u,j)}+\frac{(1-\alpha)\hat{t}(u,v)}{\sum_{j\in U}\hat{t}(u,j)},
\end{equation}
where $0\leq\alpha\leq1$ denotes the linear weighting coefficient.
The standard CF algorithm then corresponds to the case in which $\alpha=1$.
Through the analysis on Eqns.(\ref{eqn:f1}) and (\ref{eqn:f2}), we see that the standard CF framework totally neglects the impact of
the user trust. We show that in this paper, by taking into account of user trust, the \emph{iTrace} algorithm can provide
more accurate recommendations compared with
the standard CF method. Through evaluation on a public dataset, we also demonstrate that Eqn.(\ref{eqn:f1}) is
preferable to Eqn.(\ref{eqn:f2}) in Sec.\ref{sec:exp}.

The design of the implicit trust inference procedure, which is involved at Step 3 as shown above, creates a difference
between the proposed \emph{iTrace} algorithm and the other existing
TaCF methods. The connections to related work in the literature are presented in Sec.\ref{subsec:connect}.
We describe in detail the implicit trust inference procedure in the next subsection.
\subsection{Implicit Trust Inference}\label{subsec:trust_inf}
This module takes as input an $N\times N$ sparse explicit trust matrix $T_e$, and exploits trust propagation in order to predict,
for every user, how much she could trust every other user. In this way, it outputs an estimated trust matrix $\hat{T}$,
the $(i,j)$th cell $\hat{t}(i,j)$ (if present) of which represents how much the $i$th user trusts the $j$th user.
The input matrix $T_e$ has a very limited number of
cells that take value 1 and all the other cells are empty with missing values.
If the $(i,j)$th cell of $T_e$, denoted by $t_e(i,j)$, takes value 1, it means that user $i$ has expressed a trust statement that she trusts user $j$.
It is worth noting that, in practical applications, the available trust data, represented as matrix $T_e$ here,
would always be very sparse. This is so because no user can reasonably interact with every other user and then issue a trust statement about them.
\subsubsection{Four categories of trust information flow patterns}\label{sec:category}
\begin{figure}[t]
\centering
\includegraphics[width=2.3in,height=3in]{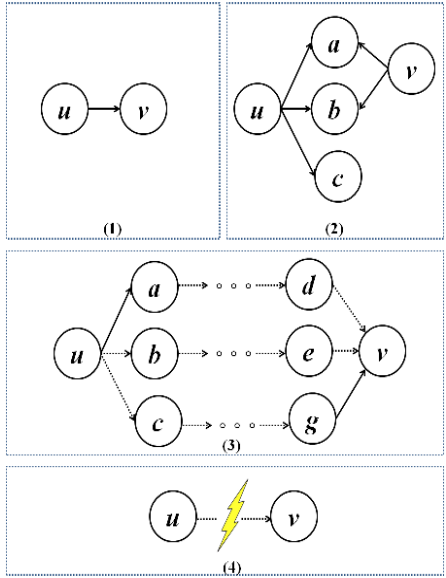}
\caption{4 typical example cases under consideration for predicting user $u$'s trust on user $v$.
A solid line with an arrow pointing from $i$ to $j$ is associated with the event that
$t_e(i,j)$ takes value 1. A dotted line with an arrow pointing from $i$ to $j$ indicates a missing value of $t_e(i,j)$ and that there exists
an implicit trust of $i$ on $j$, which can be inferred from $T_e$. The lightning symbol in the 4th sub-figure
indicates a cut-off of the trust information flow.}\label{fig:trust_pred_cases}
\end{figure}

We categorize the patterns of the trust information flow from user $u$ to $v$ into 4 complementary classes. For each class, we show a typical example
in Fig.\ref{fig:trust_pred_cases}.
The 1st sub-figure corresponds to the case in which $t_e(u,v)=1$.
If $t_e(i,j)=1$, then $j$ is called an explicit trustee of $i$. The 2nd sub-figure exemplifies the case in which
$v$ is not an explicit trustee of $u$ but $u$ and $v$ have common explicit trustee(s).
The 3rd sub-figure exemplifies the case in which $v$ is not an explicit trustee of $u$,
$u$ and $v$ have no common explicit trustee but there is at least one trust propagation path from $u$ to $v$.
A trust propagation path from $i$ to $j$ is defined by a series of user
pairs $\{p_m, p_{m+1}\}, m=1, \ldots, M-1, M\in\mathbb{N}$, which satisfies $p_1=i, p_M=j$, and $\hat{t}(p_m, p_{m+1})>0$, $\forall m\in\{1,\ldots,M-1\}$.
This model indicates that, if user $i$ trusts another user $k$ to some extent and user $k$ trusts user $j$ to some extent, then
there will be a trust propagation path from $i$ to $j$. Note that this model conforms to the transitivity property of the concept of
trust \cite{falcone2012trust,papagelis2005alleviating,massa2004trust}.
The 4th sub-figure is associated with the case that there is no trust propagation path from $u$ to $v$.
\subsubsection{Trust inference procedures}
\begin{figure}[t]
\centering
\includegraphics[width=3.3in,height=1.2in]{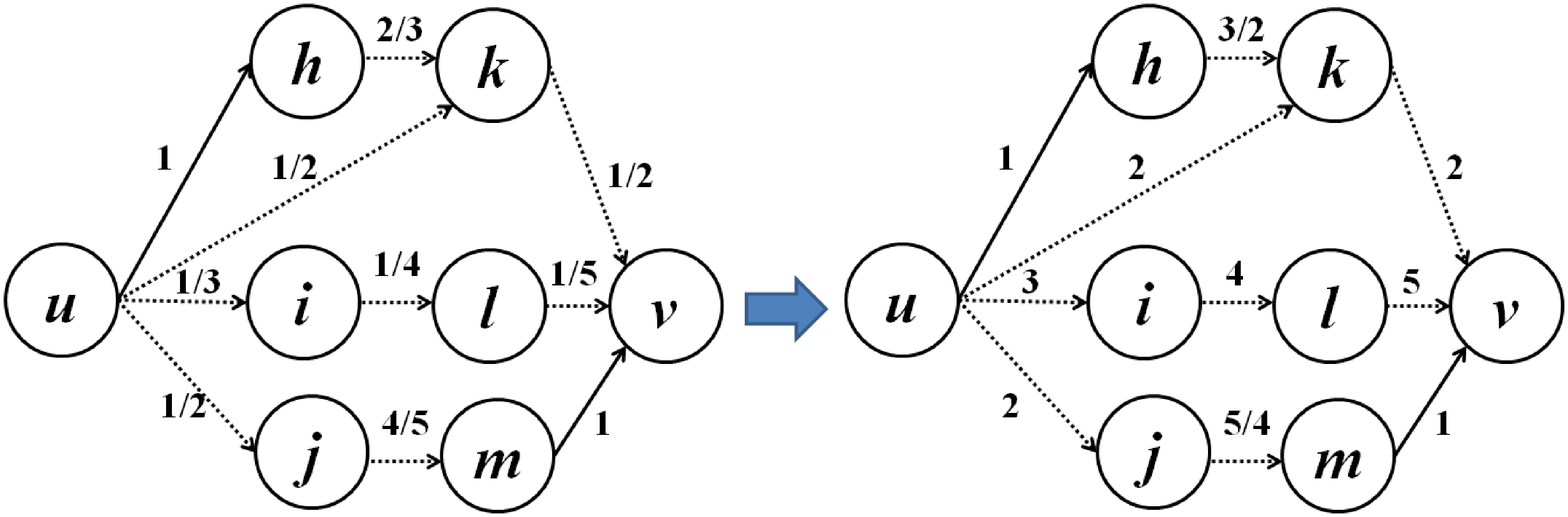}
\caption{An example show of the transformation from a weighted directed graph (the left panel) to its reciprocal counterpart graph (the right panel).
The edge weight in the right graph is the reciprocal of the weight in the left graph. }\label{fig:inverse_trust}
\end{figure}

First we initialize $\hat{T}$ to be a zero matrix.
Given a pair of users, $u$ and $v$, we first determine which one of the 4 categories presented above
the pattern of the trust information flow from $u$ to $v$ belongs to.
If it belongs to the 1st category, set $\hat{t}(u,v)=1$.
If it belongs to the 2nd category, we estimate $\hat{t}(u,v)$ as follows
\begin{equation}
\hat{t}(u,v)=\frac{|S(u)\bigcap S(v)|}{|S(u)\bigcup S(v)|},
\end{equation}
where $S(i)$ denotes the set of $i$'s explicit trustees. For the case shown in the 2nd sub-figure of Fig.\ref{fig:trust_pred_cases}, we then have
$\hat{t}(u,v)=2/3$. Now we focus on the case in which the pattern of the trust information flow from $u$ to $v$ belongs to the 3rd category.
We consider every user pair $\{i,j\}$ that belongs to any one of the aforementioned two categories and
estimate $\hat{t}(i,j)$ correspondingly.
Then we exploit trust propagation over the trust network defined by $\hat{T}$ to estimate $\hat{t}$ for user pairs associated with the 3rd category.
We treat the trust matrix $\hat{T}$ as a weighted directed graph $G$, in which the nodes denote the users and the weight of an edge denotes the
trust of the starting vertex on the end vertex.
For a pair of users, say $u$ and $v$, there may be multiple paths originating from $u$ and ending at $v$, as exemplified in
the 3rd sub-figure of Fig.\ref{fig:trust_pred_cases}.
To compute $\hat{t}(u,v)$, we first build up a reciprocal trust matrix $\hat{T}_{r}$,
whose $(i,j)$th cell $\hat{t}_r(i,j)=1/\hat{t}(i,j)$ if $\hat{t}(i,j)>0$; otherwise, set $\hat{t}_r(i,j)=\infty$.
We treat the reciprocal trust matrix as a weighted directed graph $G_r$. An exemplary show of the transformation from $G$ to $G_r$ is presented in
Fig.\ref{fig:inverse_trust}. Then we consider a shortest path problem \cite{ahuja1990faster}, which aims to find the shortest path from $u$ to $v$,
denoted by $SP_r(u,v)$, in $G_r$. The Dijkstra's algorithm \cite{chen2003dijkstra} is employed here to find $SP_r(u,v)$.
Then we set $\hat{t}(u,v)$ as follows
\begin{equation}\label{eqn:trust_sp}
\hat{t}(u,v)=\frac{1}{M\times L(SP_r(u,v))},
\end{equation}
where $L(\cdot)$ and $M$ denote the length of a path and the number of nodes
except the starting node included in
the shortest path, respectively. For the sake of clarity, consider the case shown in the right graph of Fig.\ref{fig:inverse_trust},
for which the shortest path from $u$ to $v$, $SP_r(u,v)$, is $u\rightarrow k\rightarrow v$,
$L(SP_r(u,v))=\hat{t}_r(u,k)+\hat{t}_r(k,v)=4$, $M=2$ and thus $\hat{t}(u,v)=1/8$.

If the pattern of the trust information flow from $u$ to $v$ does not belong to any of the above mentioned categories,
it then must belong to the 4th category, for which we set $\hat{t}(u,v)=0$.

It is worth noting that, given $\hat{T}$, the computation of $\hat{t}(u,v)$ for all cases included in the 1st, 3rd and 4th categories
can be unified by a single formula as follows
\begin{equation}\label{eqn:unify_trust}
\hat{t}(u,v)=\frac{1}{M\times\sum_{m=1}^{M-1}\frac{1}{\hat{t}(p_m,p_{m+1})}},
\end{equation}
where $p_1=u, p_M=v$ and $p_1\rightarrow p_2\rightarrow\ldots\rightarrow p_M$ is the shortest path in $G_r$ from $u$ to $v$.
The 1st category defined in Sec.\ref{sec:category} is associated with cases in which $M=2$ and $\hat{t}(p_1,p_2)=1$.
The 4th category corresponds to cases in which, $\exists m\in\{1,\ldots,M-1\}$, $\hat{t}(p_m,p_{m+1})=0$.
With the aid of Eqn.(\ref{eqn:unify_trust}), we can infer that, provided all the other conditions are the same,
the bigger the value of $M$ is or the smaller the value of $\hat{t}(p_m,p_{m+1})$ is, the smaller the value of $\hat{t}(u,v)$ will be,
and vice versa. The above effect is consistent with our intuitive understanding of the property of transitivity in the trust type relationships
between a pair of users.
\subsection{Connections to related work}\label{subsec:connect}
The \emph{iTrace} algorithm proposed here finds connections to several existent TaCF methods in the literature.
The algorithm architecture of \emph{iTrace} falls within a generic TaCF framework presented in \cite{massa2004trust},
while the implicit inference procedure of \emph{iTrace} presented here is unique.
In a variety of existent TaCF methods \cite{abdul2000supporting,lathia2008trust,papagelis2005alleviating,hwang2007using,pitsilis2004model},
the trust score is derived from the user rating data. To this regard, trust inference and the computation of user similarity are performed based on
exactly the same information source. In contrast with the aforementioned work, the \emph{iTrace} algorithm employs not only the
user rating data but also data other than user rating, namely the explicit user trust data.
Since the explicit trust data and the rating data are processed independently, the \emph{iTrace} has the advantage of 
making full use of two complementary views in rating prediction. Furthermore, since the trust inference 
procedure can be performed offline prior to calculation of user similarity, the computation time of the \emph{iTrace} for online rating predictions
is similar to traditional CF methods. 

The \emph{iTrace} algorithm also finds connections to our previous work on trust modeling in the context of wireless sensor
networks \cite{liu2015toward,wang2017online,liu2016state}. Although the same term
trust is used, its physical meaning is different. In \emph{iTrace}, the term trust represents a classical social relationship among users,
while in \cite{liu2015toward,wang2017online,liu2016state},
it is an artificially designed concept related to abnormal sensory behaviors caused by sensor faults.

To our knowledge,
the most similar work to our algorithm is a trust based CF method presented in \cite{chen2016trust}, which has come to our attention only recently.
In contrast with \cite{chen2016trust}, we provide a new and more efficient way for readers to understand the
shortest path based formulation of the trust inference problem by identifying four categories of trust information flow patterns and unifying three of them
by a single formula, namely Eqn.(\ref{eqn:unify_trust}). In addition, we consider two different
ways, namely IW and LW as specified by Eqns.(\ref{eqn:f1}) and (\ref{eqn:f2}), respectively, for fusion of trust and similarity; while in
\cite{chen2016trust}, only one way, i.e., LW, is considered.
Further, the \emph{iTrace} algorithm leverages the property of trust value attenuation in the trust propagation process by adding a penalization item $M$ to
the denominator of Eqn.(\ref{eqn:trust_sp}), while the method in \cite{chen2016trust} does not take into account of such attenuation effect.
Finally, through a public open dataset, we demonstrate that our \emph{iTrace} algorithm outperforms the method proposed in \cite{chen2016trust}.
\section{Performance Evaluation}\label{sec:exp}
In this section, we present experimental results, which show that the proposed \emph{iTrace} algorithm outperforms existing competitor methods.
We conducted empirical performance evaluations on the public dataset Filmtrust \cite{golbeck2006filmtrust}.
\subsection{About the dataset}
The Filmtrust dataset consists of a $N\times M$ rating matrix $R$ and an $N\times N$ explicit trust matrix $T_e$,
associated with $N=1508$ users and $M=2071$ movie items.
The $(i,j)$th cell of $R$ is filled with user $i$'s rating on item $j$ if it exists; otherwise it is empty.
A total number of 35416 ratings are included in $R$, whose values are between 0.5 and 4; and the empty cells of $R$ are
missing values to be predicted.
The matrix $T_e$ is sparse in that only 1642 cells of it are filled with value 1 associated with a set of trust statements and
the other cells are empty, corresponding to missing values to be predicted.
\subsection{Experiment setting}
The comparison methods include the traditional CF algorithm, which uses the Pearson correlation as the similarity measure,
an explicit trust based TaCF (called E-TaCF for short in what follows) and a Dijkstra's algorithm based TaCF
(termed D-TaCF for short in what follows) proposed in \cite{chen2016trust}.
The traditional CF is included here as a baseline for performance comparison.
The E-TaCF algorithm can be regarded as a simplified version of \emph{iTrace} that discards the whole implicit trust inference procedure,
namely, it sets $\hat{T}$ straightforward to be $Te$ prior to the calculation of Eqn.(\ref{eqn:tacf}). We consider two types of E-TaCF, E-TaCF-I and
E-TaCF-II, corresponding to the usage of Eqn.(\ref{eqn:f1}) and of Eqn.(\ref{eqn:f2}), respectively, for fusion of similarity and trust.
The missing values in $T_e$ are filled with 0 when performing E-TaCF.
This E-TaCF algorithm is included here in order to demonstrate the value of the proposed implicit
trust procedure.
The D-TaCF is involved here as it is the most similar method in the literature to our \emph{iTrace} algorithm.
For \emph{iTrace} and E-TaCF, we consider the IW and LW weighting mechanisms both, corresponding to
Eqns.(\ref{eqn:f1}) and (\ref{eqn:f2}), respectively, and the aim is to investigate which one is better for use.
Apart from Eqn.(\ref{eqn:trust_sp}), we also considered another way to
calculate $\hat{t}(u,v)$ by
\begin{equation}\label{eqn:trust_sp2}
\hat{t}(u,v)=1/L(SP_r(u,v)).
\end{equation}
The purpose is to demonstrate that taking account of the attenuation feature of trust via Eqn.(\ref{eqn:trust_sp})
is beneficial for improving accuracy in recommendations.
To summarize, we considered in total 4 types of \emph{iTrace} as shown in Table \ref{Table:types}.
\begin{table}[h!]\small
\caption{4 types of the \emph{iTrace} algorithm under consideration}\label{Table:types}
\begin {center}
\begin{tabular}{c||c|c|c|c }
\hline  & \emph{iTrace}-I & \emph{iTrace}-II & \emph{iTrace}-III & \emph{iTrace}-IV\\
\hline choice for $f$ & Eqn.(\ref{eqn:f1}) & Eqn.(\ref{eqn:f1}) & Eqn.(\ref{eqn:f2}) & Eqn.(\ref{eqn:f2}) \\
\hline choice for $\hat{t}(u,v)$  & Eqn.(\ref{eqn:trust_sp}) & Eqn.(\ref{eqn:trust_sp2}) & Eqn.(\ref{eqn:trust_sp}) & Eqn.(\ref{eqn:trust_sp2}) \\
\hline
\end{tabular}
\end {center}
\end{table}
\subsection{Experiment results}
\begin{figure}[t]
\centering
\includegraphics[width=3.7in,height=2.3in]{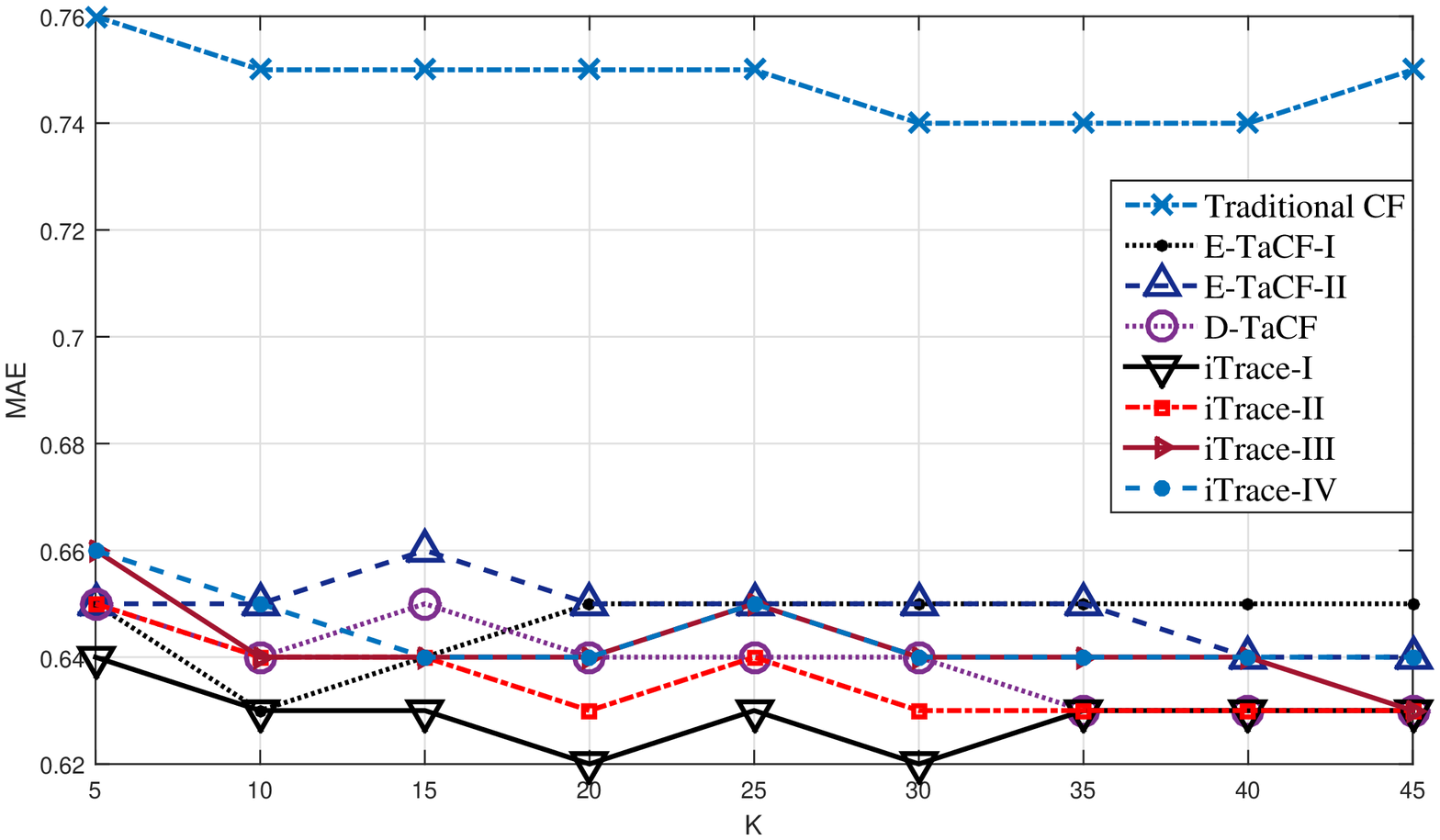}
\caption{Averaged MAE per rating prediction.}\label{fig:mae_comp}
\end{figure}

In our experiment, the number of voting users $K$ takes values in $\{5,10,15,20,25,30,35,40,45\}$, and for every $K$ value, a cross validation type
test for every comparison method is performed.
We partition the sample of rating data into two complementary subsets, perform the user similarity analysis on one subset, which
occupies 80$\%$ of the whole dataset, hiding the other 20$\%$ ratings and trying to predict them. The predicted rating is then compared with the real
rating and the difference (in absolute value) is the prediction error. The mean absolute error (MAE) is adopted
as the performance measure.
To reduce variability, we perform 5 rounds of the above operations using different partitions for each algorithm,
and the prediction results are averaged over the rounds. For the sake of fairness in comparison,
we try different $\alpha$ values and then select the optimal value $0.3$ for use for \emph{iTrace}-III and \emph{iTrace}-IV.
The comparison result, in terms of averaged MAE per rating prediction, is presented in Fig.\ref{fig:mae_comp}.
It is shown that trust based methods outperform the traditional CF significantly and that
the \emph{iTrace}-I algorithm beats all the other competitors.
It also indicates that Eqn.(\ref{eqn:f1}) is preferable to Eqn.(\ref{eqn:f2}) and Eqn.(\ref{eqn:trust_sp}) is preferable to
Eqn.(\ref{eqn:trust_sp2}) for use in implementing \emph{iTrace}. Note that, since the implicit trust inference procedure is performed offline,
the computation time of \emph{iTrace} for online rating prediction is similar to the traditional CF method.
\section{Conclusions}\label{sec:con}
In this paper, we proposed an improved CF algorithm termed \emph{iTrace},
which is featured by an embedded powerful implicit trust inference method.
This method can estimate the implicit trust relationship between a pair of users based on
available but very limited explicit trust information among users.
The result from an extensive experiment on a public dataset demonstrates the superiority of our algorithm to existent competitors.
Future work lies in using the proposed technique to analyze more datasets. How to model and employ
more social interactions among users and the temporal dynamics in the users' rating behaviors to improve CF is also a
promising topic for future investigation.
\bibliographystyle{IEEEtran}
\bibliography{mybibfile}
\end{document}